\documentclass[final,1p,11pt]{elsarticle}
\pdfoutput=1
\usepackage{latexsym}
\usepackage{amsmath}
\usepackage{multirow}
\usepackage{url}
\usepackage{graphicx}
\usepackage[ruled,vlined]{algorithm2e}
\usepackage{amssymb}
\usepackage{amsfonts}
\usepackage{todonotes}
\usepackage{fancyvrb}
\usepackage{microtype}
\usepackage{mathpazo}
\usepackage{tikz}
\usetikzlibrary{bayesnet}
\usepackage[colorlinks,urlcolor=blue]{hyperref}

\title{jLDADMM: \\ A Java package for the LDA and DMM topic models}

\author{Dat Quoc Nguyen}

\address{School of Computing and Information Systems \\
 The University of Melbourne, Australia\\
\tt{{dqnguyen@unimelb.edu.au}}}

\begin{document}

\begin{abstract}
\textbf{Abstract:} In this technical report, we present jLDADMM---an  easy-to-use Java toolkit for  conventional topic models.  jLDADMM is released to provide alternatives for topic modeling on normal or short texts. It provides implementations of the Latent Dirichlet Allocation topic model and the one-topic-per-document Dirichlet Multinomial Mixture model (i.e. mixture of unigrams), using collapsed Gibbs sampling. In addition, jLDADMM supplies a document clustering evaluation to compare topic models. jLDADMM is  open-source and available to download at: \url{https://github.com/datquocnguyen/jLDADMM}

\end{abstract}
\maketitle

\section{Introduction}

\noindent  Topic modeling algorithms  are 
statistical methodologies ``for analyzing documents, where a document is viewed as a collection of words, and the words in the document are viewed as being generated by an underlying set of topics'' \citep{Jordan255}.\footnote{In fact,  topic models are also used for other kinds of data  \citep{Blei2012}. However,  in this report we discuss topic modeling in the context of text analysis.}  The 
probabilistic topic model  Latent Dirichlet Allocation (LDA) \cite{Blei2003} is the most widely used model to discover latent topics in document collections. However, as shown in \citet{Tang:2014}, LDA obtains poor performance when the data presents extreme
properties (e.g., very short or very few documents). That is, applying topic models to short documents, such as Tweets or instant messages, is challenging because of data sparsity and the limited contexts in such texts. One approach is to combine short texts into long pseudo-documents before training LDA \citep{Hong2010D,Weng2010,Mehrotra2013,Bicalho:2017:GFE:3062405.3062584}. Another approach is to assume that there is only one topic per document \citep{Nigam2000,Zhao2011,Yin2014,jmir.6045}, such as in the mixture of unigrams Dirichlet Multinomial Mixture (DMM) model \citep{Yin2014}. 

We present in this  technical report jLDADMM---a Java package for the LDA and DMM topic models. 
 jLDADMM is released to provide alternative choices for topic modeling on normal or short texts.  
jLDADMM provides implementations of the LDA topic model \cite{Blei2003} and the one-topic-per-document DMM  \cite{Nigam2000}, using  the collapsed Gibbs sampling algorithms for inference as described in \cite{GriffithsS2004} and \cite{Yin2014}, respectively. Furthermore, jLDADMM supplies a document clustering evaluation to compare topic models, using two common metrics of Purity and normalized mutual information (NMI) \cite{ManningRS2008}.  

Our design goal is to make jLDADMM simple to
setup and run. All jLDADMM components
are packaged into a single  file {\tt .jar}. Therefore,  users
do not have to install external dependencies.
Users can run jLDADMM from either
the command-line or the Java API. The next sections will detail the usage of jLDADMM in command line, while examples of using the API are available  at \url{https://github.com/datquocnguyen/jLDADMM}.

 Please \textbf{cite} jLDADMM when it is used to produce published results or incorporated into other software. 
Bug reports, comments and suggestions about jLDADMM are highly appreciated. As a free open-source package, jLDADMM is distributed on an "AS IS" BASIS, WITHOUT WARRANTIES OR CONDITIONS OF ANY KIND, either express or implied.

\section{Using jLDADMM for topic modeling}

\noindent This section describes the usage of jLDADMM in command line or terminal, using a pre-compiled file named {\tt jLDADMM.jar}. Here, it is supposed that Java is already set to run in command line or terminal (e.g. adding Java to the environment variable {\tt path} in Windows OS).

Users can find the pre-compiled file \texttt{jLDADMM.jar}  and source codes in folders \texttt{jar} and \texttt{src}, respectively. Users can also recompile the source codes by simply running \texttt{ant} (it is also expected that \texttt{ant} is already installed). In addition,  users can find input examples in folder   \texttt{test} .

\textbf{File format of input corpus: } Similar to file \texttt{corpus.txt}  in    folder \texttt{test}, jLDADMM assumes that \textit{each line in the input corpus file represents a document}. Here, a document is a sequence of  words/tokens separated by white space characters. The users should preprocess the input corpus before training the LDA or DMM topic models, for example: down-casing, removing non-alphabetic characters and stop-words, removing words shorter than 3 characters and words appearing less than a certain times.   

\medskip

\underline{Now, we can train LDA or DMM by executing:}

\vspace{5pt}

\noindent \texttt{\$ java [-Xmx1G] -jar jar/jLDADMM.jar -model <LDA\_or\_DMM> -corpus \\ <Input\_corpus\_file\_path> [-ntopics <int>] [-alpha <double>] [-beta <double>] [-niters <int>] [-twords <int>] [-name <String>] [-sstep <int>]}

\vspace{5pt}

\noindent where parameters in [ ] are optional. 

\begin{itemize}

\item \texttt{-model}: Specify the topic model LDA or DMM

\item \texttt{-corpus}: Specify the path to the input  corpus file.

\item \texttt{-ntopics <int>}: Specify the number of topics. The default value is 20.

\item \texttt{-alpha <double>}: Specify the hyper-parameter $\alpha$. Following \cite{Yin2014,LuMZ2011}, the default $\alpha$ value is 0.1.  

\item \texttt{-beta <double>}: Specify the hyper-parameter $\beta$. The default $\beta$ value is 0.01 which is a common setting in the literature \cite{GriffithsS2004}. Following \cite{Yin2014}, the users may consider to the $\beta$ value at 0.1 for short texts. 

\item \texttt{-niters <int>}: Specify the number of Gibbs sampling iterations. The default value is 2000.

\item \texttt{-twords <int>}: Specify the number of the most probable topical words. The default value is 20.

\item \texttt{-name <String>}: Specify a name to the topic modeling experiment. The default value is \textit{``model''}.

\item \texttt{-sstep <int>}: Specify a step to save the sampling outputs. The default value is 0 (i.e. only saving the output from the last sample).

\end{itemize}

\textbf{Examples:}

\vspace{5pt}

\noindent {\texttt{\$ java -jar jar/jLDADMM.jar -model LDA -corpus test/corpus.txt -name testLDA}}

\vspace{5pt}

 The output files are saved in the same folder containing the input  corpus file, in this case: the  folder \texttt{test}. We have output files of \texttt{testLDA.theta}, \texttt{testLDA.phi}, \texttt{testLDA.topWords}, \texttt{testLDA.topicAssignments} and \texttt{testLDA.paras}, referring to the document-to-topic distributions, topic-to-word distributions, top topical words, topic assignments and model parameters, respectively.  

Similarly, we perform:

\vspace{5pt}
\noindent { \texttt{\$ java -jar jar/jLDADMM.jar -model DMM -corpus test/corpus.txt -beta 0.1 -name testDMM}}

\vspace{5pt}

Output files  \texttt{testDMM.theta}, \texttt{testDMM.phi}, \texttt{testDMM.topWords},  \texttt{testDMM.topicAssignments} and \texttt{testDMM.paras} are also in   folder \texttt{test}.

\section{Topic inference on new/unseen corpus}
\noindent  To infer topics on a new/unseen corpus using a pre-trained LDA/DMM topic model, we perform:

\texttt{\$ java -jar jar/jLDADMM.jar -model <LDAinf\_or\_DMMinf> -paras \\ <Hyperparameter\_file\_path> -corpus <Unseen\_corpus\_file\_path> [-niters <int>] [-twords <int>] [-name <String>] [-sstep <int>]}

\begin{itemize}
\item  \texttt{-paras}: Specify the path to the hyper-parameter file produced by the pre-trained LDA/DMM topic model.
\end{itemize}

\textbf{Examples:}

\vspace{5pt}

\noindent  \texttt{\$ java -jar jar/jLDADMM.jar -model LDAinf -paras test/testLDA.paras -corpus test/unseenTest.txt -niters 100 -name testLDAinf}

\vspace{5pt}
\noindent  \texttt{\$ java -jar jar/jLDADMM.jar -model DMMinf -paras test/testDMM.paras -corpus test/unseenTest.txt -niters 100 -name testDMMinf}

\section{Using jLDADMM for document clustering evaluation}

\noindent We treat each topic as a cluster, and we assign every document the topic with the highest probability given the document \cite{LuMZ2011}. 
To get the Purity and NMI clustering scores, we perform:

\vspace{5pt}
\noindent  \texttt{\$ java -jar jar/jLDADMM.jar -model Eval -label <Golden\_label\_file\_path> -dir  <Directory\_path> -prob <Document-topic-prob/Suffix>}

\begin{itemize}
\item \texttt{-label}: Specify the path to the ground truth label file. Each line in this label file contains the golden label of the corresponding document in the input  corpus. See files \texttt{corpus.LABEL} and \texttt{corpus.txt}  in   folder \texttt{test}.

\item \texttt{-dir}: Specify the path to the directory containing document-to-topic distribution files. 

\item \texttt{-prob}: Specify a document-to-topic distribution file or a group of document-to-topic distribution files in the specified directory.

\end{itemize}

\textbf{Examples:}

\vspace{5pt}
\noindent {\texttt{\$ java -jar jar/jLDADMM.jar -model Eval -label test/corpus.LABEL -dir   test -prob testLDA.theta}

\vspace{5pt}
\noindent  {\texttt{\$ java -jar jar/jLDADMM.jar -model Eval -label test/corpus.LABEL -dir  test -prob testDMM.theta}
\vspace{5pt}

The above commands will produce the clustering scores for files \texttt{testLDA.theta} and \texttt{testDMM.theta}  in   folder \texttt{test}, separately.  

The following command:

\vspace{5pt}
\noindent  \texttt{\$ java -jar jar/jLDADMM.jar -model Eval -label test/corpus.LABEL -dir   test  -prob theta}
 \vspace{5pt}

\noindent will produce the clustering scores for all  document-to-topic distribution files with their names ending in \texttt{theta}. In this case, they are \texttt{testLDA.theta} and \texttt{testDMM.theta}. It also provides the \textit{mean} and \textit{standard deviation} of the  scores. 

\bigskip 

 To improve evaluation scores, the users might consider combining the LDA and DMM  models with word embeddings \cite{TACL582}, with the source code at \href{https://github.com/datquocnguyen/LFTM}{HERE}.

\bibliographystyle{elsarticle-harv}
\bibliography{References}

\end{document}